**Protective coatings for front surface silver mirrors by atomic layer deposition.**

Pavel Bulkin[1,4], Sofia Gaiaschi[2], Patrick Chapon[2], Dmitri Daineka[1], Natalya Kundikova[3,4]

*1. LPICM, UMR 7647 CNRS-Ecole Polytechnique, Institut Polytechnique de Paris, Route de Saclay, 91128 Palaiseau Cedex, France*

*2. HORIBA FRANCE S.A.S., Avenue de la Vauve, Passage Jobin Yvon, 91120 Palaiseau, France*

*3. Nonlinear Optics laboratory, Institute of Electrophysics, Ural Branch of the Russian Academy of Sciences, Yekaterinburg, 620016 Russia*

*4. Optoinformatics Department, Physics Faculty, Institute of Natural Sciences and Mathematics, South Ural State University, Chelyabinsk 454080, Russia*

ABSTRACT:

Silver is a metal which provides the highest reflectivity in the very broad wavelength range as well as the lowest polarization splitting. However, it is not very stable chemically and silver mirrors are easily damaged in a corrosive or oxidizing environment, leading first to the drastic drop in reflection followed by the complete disintegration of a silver layer. For this reason aluminum is much more in use. The problem of protection of silver layer is a very important one for number of applications, requiring the front side reflection, such as telescopes mirrors, reflective IR imaging optics, gratings, photovoltaic concentrator mirrors, etc. Atomic layer deposition (ALD) technique using trimethylaluminum (TMA) and water as precursors provides a very efficient way to protect a sensitive surface of silver from a corrosive and oxidizing environment, because ALD coatings can be deposited at rather low temperature. Moreover, ALD layer provides extremely high conformality (even when deposited over high aspect ratio features) and has high integrity, efficiently blocking foreign species diffusion to silver-overcoat interface.

In our studies we tested the efficiency of the protection of silver mirrors by ALD-deposited $Al_2O_3$ layers against oxygen plasma exposure by correlating the ellipsometric measurements with the absolute reflection measurements and Glow-Discharge Optical Emission Spectroscopy (GD-OES) data. We have found that for optimal protection the thickness of ALD deposited layer should exceed at least 15 nm (about 150 ALD cycles at 150 $^o$C), as thinner layers do not provide reliable protection of silver surface against oxygen plasma. We have also demonstrated that the deposition of 15 nm of a protective ALD-deposited $Al_2O_3$ layer does not affect the absolute reflectivity of a silver mirror in a spectral range 300 -2500 nm.

INTRODUCTION:

Silver is an ideal metal for the front surface mirrors for optics in both visible and infrared wavelength ranges. In this range silver possesses the highest reflectivity, lowest emissivity and lowest polarization splitting of all known metals [1]. Silver, however, can be oxidized very easily and requires reliable protection to be stable in the aggressive operating environment. The problem of protection of silver layers is a very important one for number of practical applications, such as telescope (ground and space-born) mirrors, reflective IR imaging optics, photovoltaic concentrator mirrors, III-V laser back-reflectors, etc. [2]. Space environment especially results in severe damage arising from the exposure to highly oxidizing environment of Low Earth Orbit (LEO) [3].

Atomic oxygen, which is the most abundant corrosion precursor on low earth orbit, is formed in the space in the process of photo dissociation of $O_2$ molecules by ultraviolet (UV) photons (< 243 nm) of solar radiation which has enough energy to break a 5.12 eV $O_2$ bond in an environment where the mean free path is sufficiently long and the probability of recombination or formation of ozone is quite small. [3]

Protecting the mirrors with coatings, in this situation, is the only option to extend the lifetime span of the silver mirrors towards reasonable values. Number of different materials was reported in this respect, from silicon nitride and silicon oxide to ultra-thin Ni layers that allow not only slowing the tarnishing of silver, but also lifting the reflection in UV region, somewhat affecting though reflection in the infrared [4]. Low temperature PECVD layers of $SiN_x$ produce multiple absorption bands in the IR region due to N-H and Si-H bonds. Various deposition techniques producing densely-packed layers can be used, such as Ion Assisted Deposition or PECVD. More common thermal evaporation and electron beam evaporation unfortunately produce a columnar layer structure that allows diffusion of foreign species from the environment to the buried interface.

Using Atomic Layer Deposition (ALD) for the protection of silver-based front surface mirrors was first reported by [5-7]. While compatible with substrates of large sizes, ALD is especially well suited for the protection of small to medium-sized front surface mirrors. ALD technology can be purely thermal or plasma-assisted (PEALD). PEALD "as is" cannot be used for a deposition of entire thickness of the coating, because active oxygen will immediately destroy silver surface. So, before using PEALD the additional protective layer was required [5]. This step is not needed in purely thermal ALD, consequently, making it the best option. The most common ALD-produced material is $Al_2O_3$. It has very large transmission window with some absorption at about 900 $cm^{-1}$ but at the typical thickness of several tens of nanometers this absorption is completely negligible. High quality $Al_2O_3$ film can be deposited at temperatures as low as 100 degree Celsius, though, still more hard and dense films are obtained at the temperatures above 150 $^{o}$C [8].

In this article we report on the study of the resistance of RF magnetron sputtered front surface silver mirrors with the ALD-deposited $Al_2O_3$ protective layers to the erosion in oxygen plasma, generated in high-density plasma (HDP) MDECR system. Accelerated aging tests in plasma systems (given the high active oxygen flux to the surface) can help to optimize protective coatings performance and are not uncommon [9]. Obvious advantage of using HDP system instead of more common RF capacitively coupled plasma systems for a simulation of space environment is a very low sheath voltage, which limits energy of the oxygen ions striking the surface to just several eVs. This is consistent with the maximal impact energy of atomic oxygen on front surface mirrors during space flights [3].

EXPERIMENTAL

Silver films with the thickness around 200 nm were sputtered onto 375 μm thick 100 mm in diameter (100) silicon wafers in Alliance Concept Dp650 RF magnetron sputtering system in the following conditions: RF-power 75 Watts using 90 mm in diameter silver target, Ar pressure of 3.5 mTorr at 250 °C (substrate holder temperature). Before loading into sputtering system the silicon wafers were dipped in 5 percent hydrofluoric acid solution in water for 30 sec in order to remove native oxide. It was found that such procedure produced films of consistently better quality. Base vacuum before the depositions was always below $5 \cdot 10^{-7}$ Torr. We have found that the conditions of sputtering are very important for mirrors to withstand subsequent ALD coating deposition and our standard conditions for silver sputtering had to be adjusted in order for Ag mirror to survive $Al_2O_3$ deposition at 150 °C. Results of silver sputtering optimization are presented in Fig. 1 and 2, which compare electron microscopy images and visual appearance of silver layers after ALD depositions.

After sputtering the mirrors were transferred into Picosun R200 Advanced deposition system where they stayed under constant $N_2$ flow for one hour in order to reach stable temperature. Thermal ALD depositions were carried out at 150 °C temperature using alternating 100 ms pulses of trimethylaluminum (TMA) and water at chamber pressure of 9 Torr with constant nitrogen purge flux of 1 SLM.

Before and after depositing the protective ALD layers of different thickness, all mirrors were measured by spectroscopic ellipsometry and reflection spectroscopy. Additionally, some samples were analyzed with Glow Discharge Optical Emission Spectroscopy (GD-OES) to establish the compositional profile of the coatings [10]. Then the mirrors were exposed to high flux of oxygen ions and radicals at room temperature in high-density ECR plasma system for 1 minute. After oxygen exposure we again studied the mirrors with reflection spectroscopy, spectroscopic ellipsometry and GD-OES in order to find out the changes in reflectivity and composition and evaluate the efficiency of a protective layer.

Optical properties of ALD-deposited $Al_2O_3$ protective layers are broadly in line with those reported for the films deposited in similar conditions (Table 1). As evidenced from Fig. 3 the deposition rate versus cycles number is pretty much linear, with minor variations at low number of cycles, what could possibly be due to the fact, that silver mirrors were exposed to air before the growth for somewhat different amount of time. Those fluctuations will be studied more carefully in a future work.

Spectroscopic ellipsometer Uvisel-1 by Horiba was used for ellipsometric data acquisition in a spectral energy range 1.0 - 4.6 eV and DeltaPsi 2 software, also by Horiba, was used for data analysis. Reflection measurements on samples were performed using Perkin-Elmer Lambda 950 system equipped with a specular reflectance attachment in the wavelength range 200-2500 nm at 8 degree incidence. For GD-OES analysis we have used GD Profiler 2 by Horiba, while SEM photos of silver surface were made in Hitachi S4800 scanning electron microscope (SEM). Oxygen plasma treatment was performed in the MDECR plasma enhanced chemical vapor deposition (PECVD) system [11] at 2 mTorr pressure using 40 sccm $O_2$ flow and 1000 Watts microwave power for typical duration of 1 min. Those conditions produce constant oxygen ions flux of about $6 \cdot 10^{15}$ ions/$cm^2$·sec onto grounded wafer (around 1 mA/$cm^2$ when measured with flat Langmuir probe). Estimated oxygen radicals flux can be close to $7 \cdot 10^{16}$ per $cm^2$·sec.

As can be noticed, reflectance of silver film in the wavelength range above 320 nm is barely affected by the presence of $Al_2O_3$ layer, while for shorter wavelength reflectance differs considerably (see Fig. 5). This spectral region (<400 nm), however, is not of interest for using silver mirrors due to the strong silver absorption feature arising from interband transition at around 320 nm [12].

After oxidation step the reflection of a non-protected mirror and mirrors with both 5 and 10 nanometer-thick coatings suffered significant deterioration (see Fig. 6). Non-protected silver layer was oxidized all the way to silicon wafer, as evidenced from the feature in reflection spectrum arising due to additional contribution from polished backside of silicon wafer around the bandgap of crystalline silicon (approximately 1100 nm). While 5 nm and 10 nm layers somewhat protected bulk of silver layer, the surface still degraded significantly, leading to big drop in reflectivity in the visible wavelength range. Apparently, only the deposition of 15 nm-thick layer of $Al_2O_3$ was enough to provide full protection against erosion of silver surface by atomic oxygen. For this sample an additional plasma exposure time (5 more minutes) also did not lead to the deterioration of reflectance. For mirrors to withstand atomic oxygen attack for sufficiently long working time on LEO such coating is judged to be adequate. If other constrains are present, like mechanical damage, thermal cycling, etc. the additional tests may be required.

Analysis by GD-OES corroborates the data obtained by reflection spectroscopy. Unprotected silver film is fully oxidized through with more or less constant concentration of oxygen all the way to the point where signal from silicon wafer appears (Figs. 7a, b). 5 nm $Al_2O_3$ film apparently slows down the oxidation but about quarter of silver layer is seriously compromised (Figs 7c, d). 15 nm layer of $Al_2O_3$ film completely protects the underlying silver film from oxidation by active oxygen (Figs 7e, f). Doubling the plasma exposure time for a silver film protected with 15 nm of alumina did not produce any difference. Silver protected by thicker $Al_2O_3$ films (>15nm) also did not show deterioration during the oxygen exposure. Interestingly, for the films with thickness of 15 nm and more, a silver emission line was not appearing at the beginning of sputtering, while for thinner films it appears simultaneously with carbon line, as carbon always present on the surface, and that is true even for non-oxidized films. We verified with ellipsometry 2D mapping with a step of 1 cm that the thickness of ALD deposited films was very uniform along the surface of the mirrors. Moreover, corrosion character did not manifest itself with a spot-like erosion. We have to note here, that ellipsometry gives an average thickness across the beam spot (around one square mm) and is not sensitive to small quantity of defects, such as tiny pinholes, but they are typically absent in ALD films.

CONCLUSIONS:

Front surface silver mirrors, depending on the way they are deposited, can be quite sensitive to the ALD process. Protective films of $Al_2O_3$ deposited onto the silver mirrors by ALD provide an effective barrier against active oxygen erosion even at 15 nm of protective film thickness and can be deposited at the temperature of 150 °C, provided silver film withstands the conditions of ALD process. It is not yet clear, why the thicknesses below 15 nm could not prevent the deterioration of a mirror surface in oxygen plasma and further work is needed to clarify the details of degradation process. GD-OES technique provides fast and accurate way to investigate the degradation of mirrors after different sorts of corrosions tests.

Funding: This research did not receive any specific grant from funding agencies in the public, commercial, or not-for-profit sectors.

Table 1. Lorentz dispersion formula parameters for dielectric function of ALD-deposited $Al_2O_3$

| Parameter | $\varepsilon_\infty$ | $\varepsilon_s$ | $\omega_t$ | $\Gamma_0$ |
|-----------|------|------|-------|-------|
| Value | 0.86 | 2.65 | 12.38 | 0.047 |

List of figure captions.

Figure 1. Scanning electron microscopy images of sputtered silver films, made in standard (left) and optimal (right) conditions.

Figure 2. Impact of ALD deposition on reflection of silver sputtered in two different conditions (see inserts). Surface of silver, deposited at non-optimal conditions becomes rough and highly-scattering after the ALD process.

Figure 3. Dependence of $Al_2O_3$ layer thickness on the number of ALD cycles.

Figure 4. Ellipsometric spectra and fit for 600 cycles $Al_2O_3$ film on silver (thickness is 63 nm, material properties are given in Table 1).

Figure 5. Reflection spectra of silver mirrors before plasma oxidation measured at 8 degree incidence angle.

Figure 6. Reflection spectra of silver mirrors after plasma oxidation measured at 8 degree incidence angle.

Figure 7. GD-OES spectra of silver mirrors before and after the exposure to $O_2$ plasma.

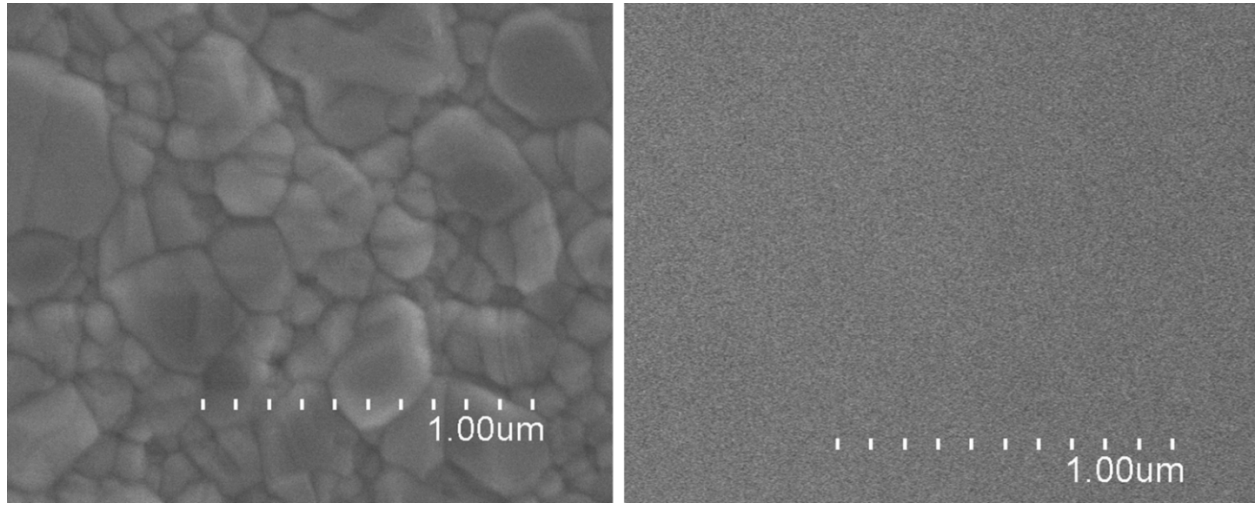

Figure 1

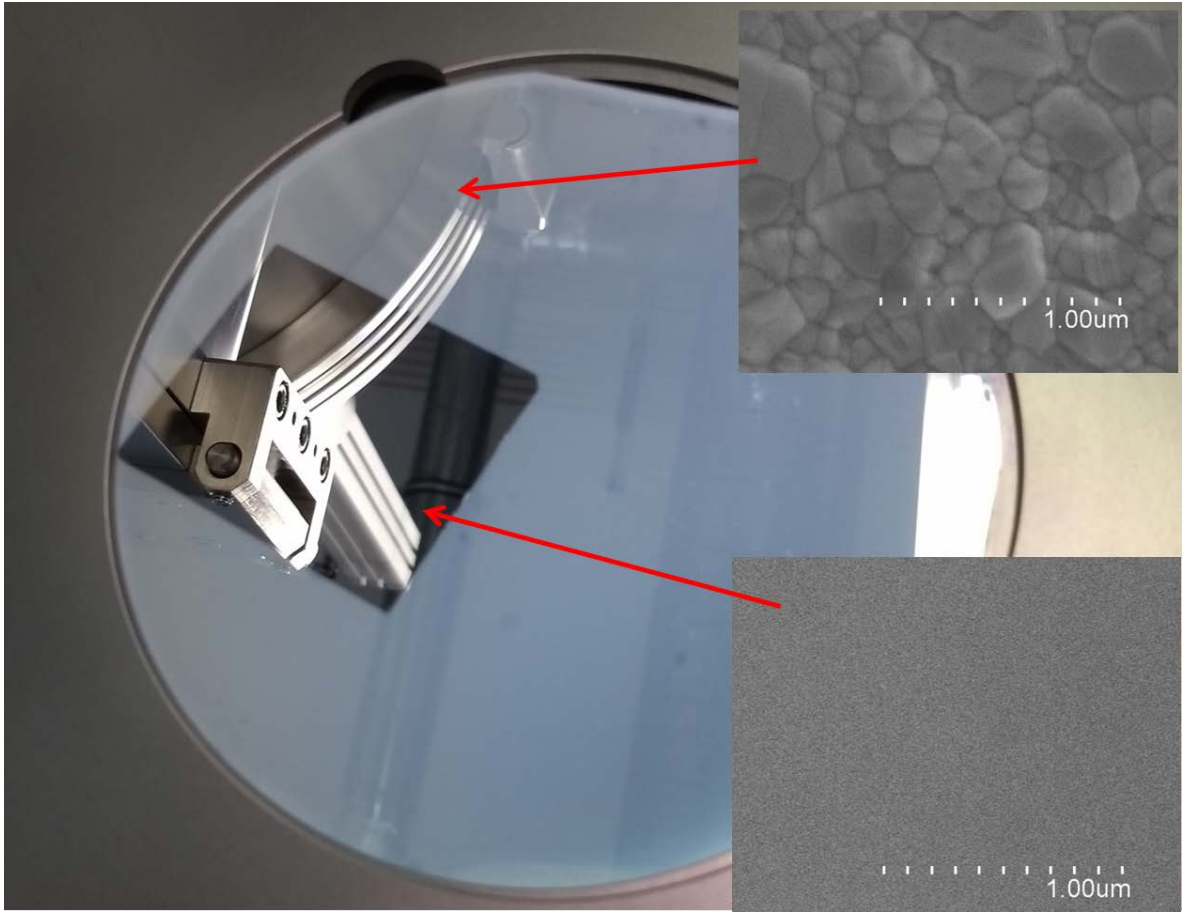

Figure 2

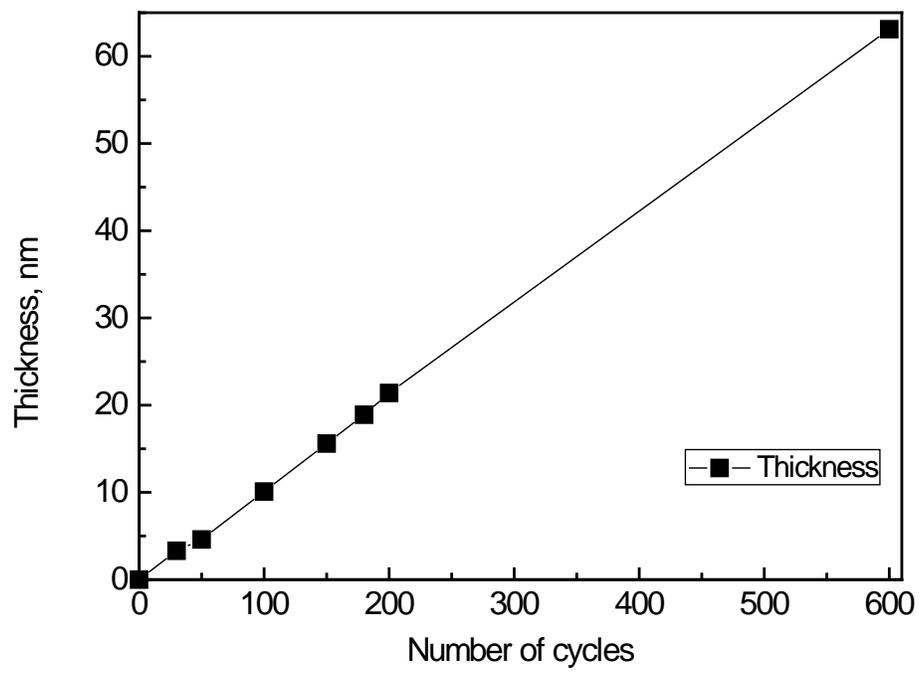

Figure 3

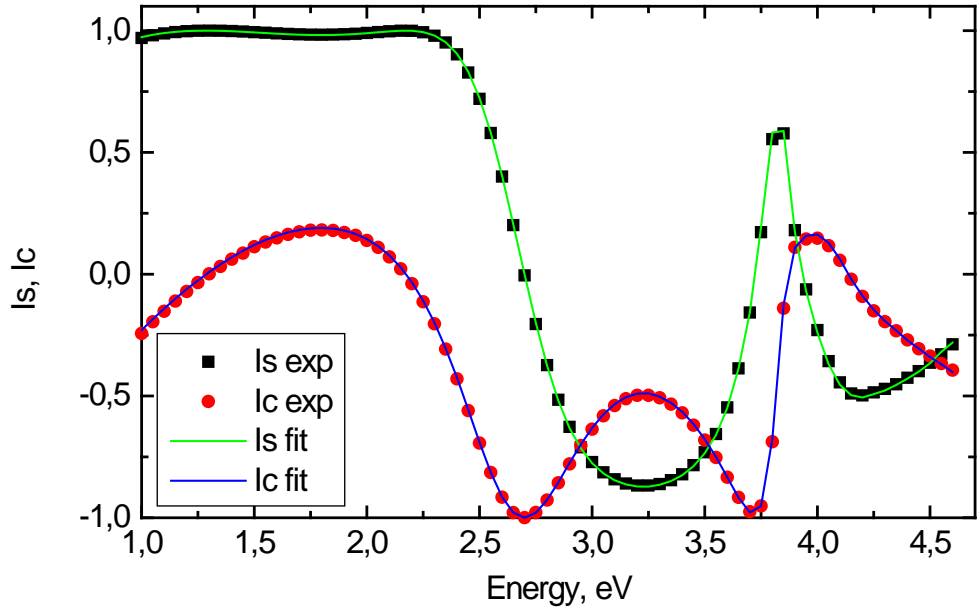

Figure 4

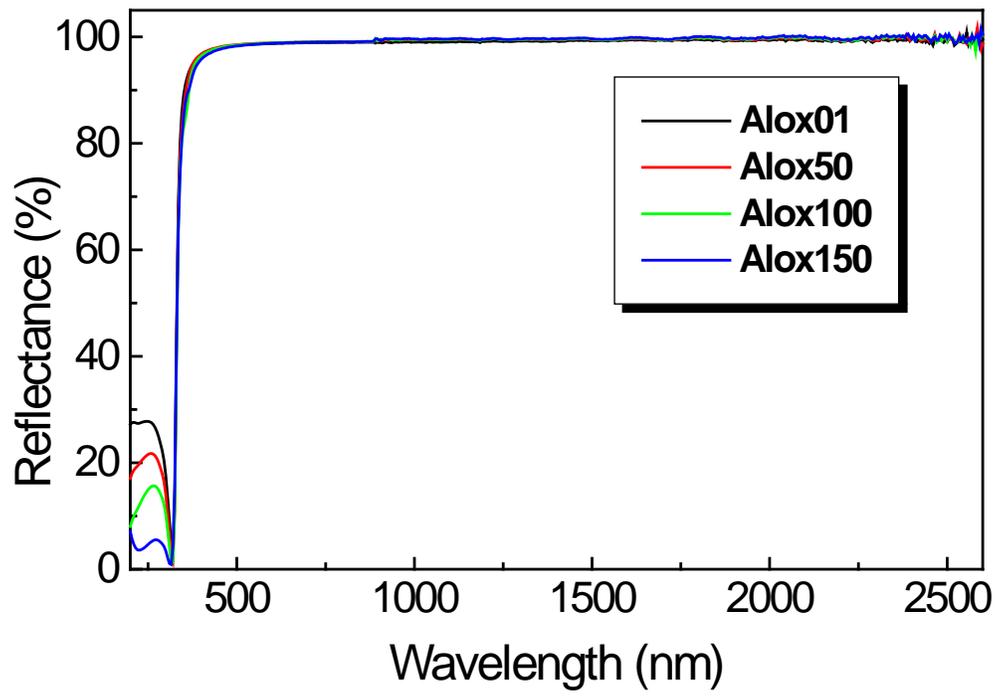

Figure 5

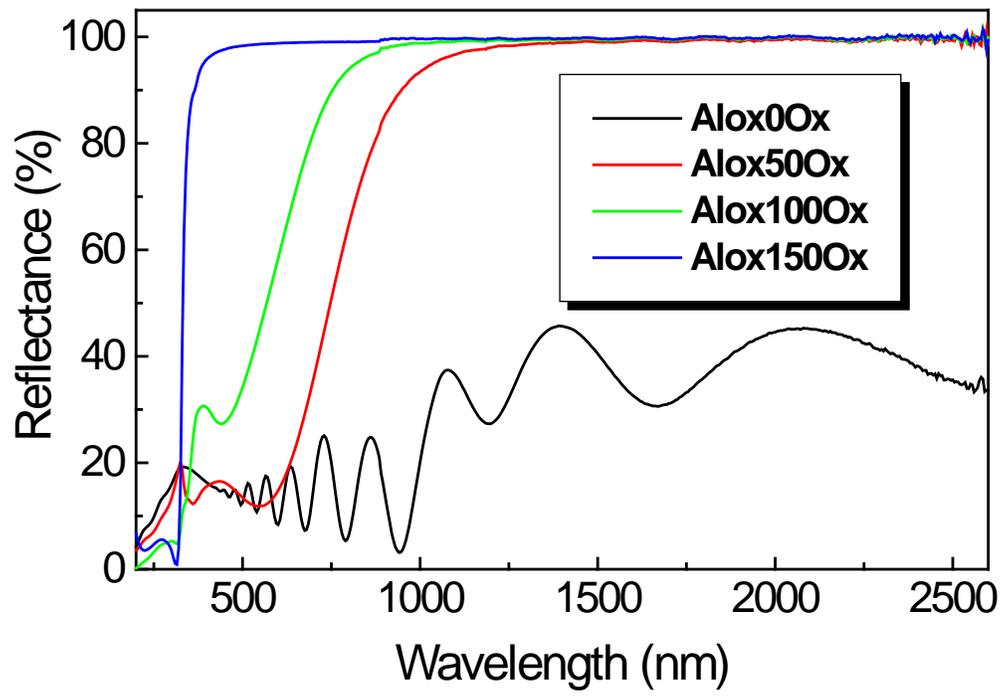

Figure 6

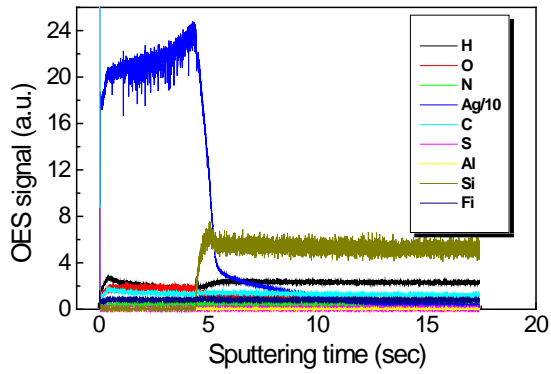

(a) Silver 0 cycles non-oxidized

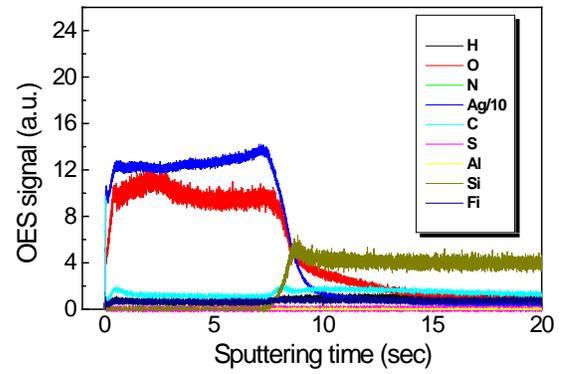

(b) 0 cycles oxidized

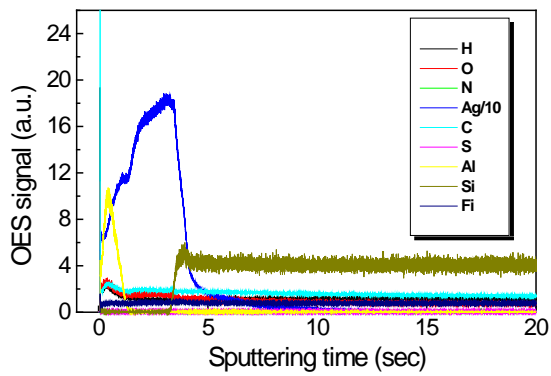

(c) 50 cycles non-oxidized

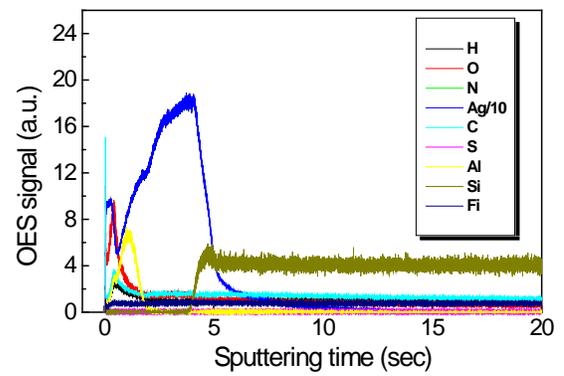

(d) 50 cycles oxidized

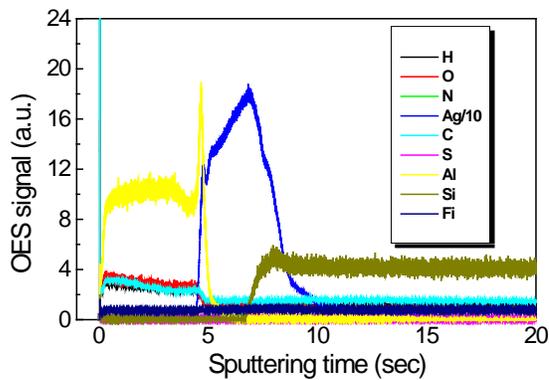

(e) 150 cycles non-oxidized

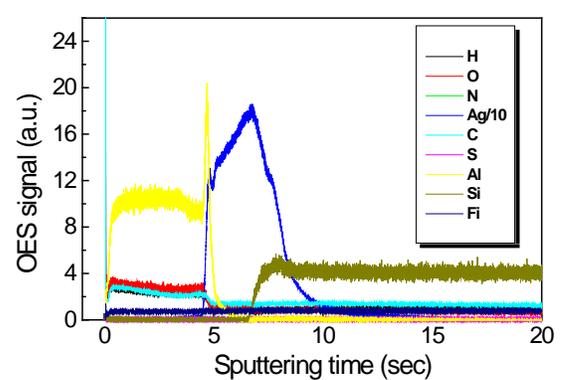

(f) 150 cycles oxidized

Figure 7